\documentclass[nosumlimits]{amsart}

\usepackage{amsmath}
\usepackage{amssymb}

\newtheorem{thm}{Theorem}[section]
\newtheorem{defin}[thm]{Definition}

\newtheorem{cor}[thm]{Corollary}

\newtheorem{pro}[thm]{Proposition}
\numberwithin{equation}{section}

 \newcommand{\into}{\rightarrow}

 \newcommand{\set}[1]{\left\{#1\right\}}
 
 \newcommand{\dotp}[2]{\left<#1,#2\right>}

 \newcommand{\qtext}[1]{\quad\text{#1}\quad}
 \newcommand{\fa}{\qtext{for all}}
 \newcommand{\bb}{\begin{equation*}}
 \newcommand{\ee}{\end{equation*}}
 \newcommand{\bp}{\begin{proof}}
 \newcommand{\ep}{\end{proof}}

\begin{document}

\title[Fields generated by moving point mass]
{Representation of fields associated with any moving point mass by
means of fundamental fields corresponding to its trajectory in the
frame of Einstein's special theory of relativity.}

\author{ Victor M. Bogdan }

\address{Department of Mathematics, McMahon Hall 207, CUA, Washington DC 20064, USA}


\email{bogdan@cua.edu}

%




%
%

\begin{abstract}
  Assume that in a Lorentzian frame is given a relativistically
admissible trajectory of a point mass. An event in such a frame
can be described by four coordinates, first three representing the
position and the last one the time of the event. Let G denote the
set of all events that do not lie on the trajectory.

  The trajectory uniquely determines on the set G a system of fields
called by the author the fundamental fields. The most important
are the following three: (1) The retarded time field, representing
the time a wave should be emitted from the trajectory to arrive at
some point of the set of events G; (2) The delayed time field,
representing the difference between the actual time of the event
and the retarded time; (3) The unit vector field representing the
direction in which the wave should be emitted.

  In the paper http://arxiv.org/abs/0909.5240 the author used
the fundamental fields to prove, that the fields of the amended
Feynman's Law satisfy the homogeneous system of Maxwell equations,
and to obtain explicit formulas for Feynman fields in terms of the
fundamental fields.

  In this note the author proves that any field on the set G of events
can be represented as a function of the three fields mentioned
above. The joint range of these three fields represents a
differentiable manifold M diffeomorphic with the set G of the
events. The manifold consists of the Cartesian product of the
space R of reals, the space of positive real numbers, and the unit
sphere in 3 dimensional Euclidean space.
\end{abstract}


\maketitle

\section{Considerations concerning trajectories}
\bigskip

Following the development in the paper of Einstein \cite{einstein2a}
we define a {\bf Lorentzian frame} to consist
of an orthogonal coordinate system in $R^3,$
having right hand orientation of axes, and equipped with a bouncing beam clock
at every point $r\in R^3$ that is synchronized with the clock located at the origin
$0\in R^3$ of the system of coordinates by means of a light beam.

In the following consideration we assume that the units of measure are selected so
that the speed of light $c=1.$

A physical event in such a frame is described by a point $(r_1,t),$
where  $r_1$ denotes a position in $R^3$ and $t\in R$ the time of the event.
Denote such a frame by $S.$ Assume that $S'$ denotes another
Lorentzian frame whose origin initially coincides with the origin of the frame $S.$
Moreover the frame  $S'$ moves as a rigid body
away from the frame of $S$ at a constant velocity.
The transformation  of coordinates of events from the frame $S$ into the frame $S'$
forms a linear transformation that preserves the quadratic form
\begin{equation*}
    |r'_1|^2-(t')^2=|r_1|^2-t^2.
\end{equation*}

A trajectory of a path of a point mass can be parameterized in several different
ways. It is important to understand which of these parameterizations depend on the
Lorentzian frame, which are invariant under Lorentzian transformations and thus
belong to Einstein's special theory of relativity, and which can be carried over
to general theory of relativity.

By {\bf geometry of Lorentz space-time} we shall understand the product space
$R^3\times R$ with the transformations of coordinates as described above.
These transformations form a group with composition of transformations as a group operation.

Though one could expand the group by adding affine transformations,
the linear transformations are sufficient for description of dynamics in physical processes
in Einstein's special theory of relativity.
Any affine orthogonal transformation can be reduced
to a linear one just by moving the origin of the coordinate system.

More general groups of transformations related to Lorentz group were studied by
several authors. For generalization of such transformations and further references
see Vogt \cite{vogt}.

Let $\alpha\mapsto y(\alpha)$ be a mapping of an interval
$I$ into $R^4,$ of class $C^2,$ in the terminology of Cartan \cite{cartan},
that is having continuous derivatives up to order $2$ on the entire interval $I.$
Assume that the mapping forms a parametric representation of a {\bf path of a point
mass} in space.

Moreover assume that the tangent vector field $y'$ consists of {\bf time-like vectors}
that is
\begin{equation}\label{y' is time like}
    y'_1(\alpha)^2+y'_2(\alpha)^2+y'_3(\alpha)^2-y'_4(\alpha)^2<0\fa \alpha \in I.
\end{equation}
As a derivative of a covariant field  with respect to a free parameter the
tensor $y'_j(\alpha)$ itself forms a covariant field over $I.$
Thus it is invariant under Lorentzian transformations and it can be carried over to
the general theory of relativity as in Dirac \cite{dirac}.
The transition from covariant to contravariant tensors is given by
means of transformation
\begin{equation*}
    y^j=g^{jk}y_k
\end{equation*}
where summation is with respect to index $k=1,2,3,4$
and the matrix $g^{jk}$ for orthogonal axes has elements on the diagonal
equal respectively $1,1,1,-1$ and non-diagonal elements are zero.

The time along the path is given by $t=y_4(\alpha),$ since
from the relation (\ref{y' is time like}) follows that that
$\frac{dt}{d\alpha}=y'_4(\alpha)>0$
for all $\alpha\in I,$ the correspondence $\alpha\mapsto t$ represents a diffeomorphism
of $I$ onto some interval $J$ and is also of class $C^2,$
that is both maps $\alpha\mapsto t=y_4(\alpha)$ and its
inverse $t\mapsto \alpha$ are of class $C^2.$

Thus in every Lorentzian frame we can represent our path in the form
\begin{equation*}
    y=(r_2(t),t)\fa t\in J,
\end{equation*}
where $t\mapsto r_2(t)$ is from some interval $J$ into $R^3$
and represents the position of the mass
as a function of time $t$ in that Lorentzian frame. Clearly this representation
is also of class $C^2$ and forms another equivalent parametric representation of the path
but this representation is frame dependent.

Most important parametrization of a path is with respect the {\bf
proper time} $s$ of the moving mass $m_0.$ The function $t\mapsto s$ is unique up to an
additive constant and can be found from the differential equation
\begin{equation}\label{proper time}
    (ds)^2=(dy_4)^2-\left((dy_1)^2+(dy_2)^2+(dy_3)^2\right)=(dt)^2-|dr_2|^2.
\end{equation}

The condition \ref{y' is time like} can be translated into
\begin{equation*}
    |\dot{r}_2(t)|=\left|\frac{dr_2(t)}{dt}\right|<1=c\fa t\in J,
\end{equation*}
that is velocity along any path of a point mass is less then the speed $c$ of light.
The notion of the proper time of a body carries over to general theory of relativity and thus
it is also invariant under Lorentzian transformations.

Maxwell established that waves in electromagnetic field in free space
propagate with velocity of light $c.$
From considerations of Einstein and Rosen \cite{einstein4} follows that even
disturbances in gravity field should propagate with velocity of light.

From results of Bogdan \cite{bogdan61} and
\cite{bogdan64}, Proposition 5.2, follows that if we consider the dynamics of
$n$ bodies interacting with each other by means of fields propagating with
velocity of light, the equations of evolution are non-anticipating differential
equations and their solutions, not only depend on the initial conditions like
in Newtonian mechanics, but also on the initial trajectory of the entire system.

Assuming for instance that in a Lorentzian frame
we are starting with $n$ bodies whose initial trajectories
$t\mapsto y_j(t),$ where $j=1,\ldots,n,$ are known and we intend to observe
the dynamics of evolution of the system for a period of time $t_1,$ and we can
a priori estimate the bound $v$ on velocities, and the bound $A$ on the accelerations
and the initial diameter $\delta$ of the system, then the length of the interval
of significance is at most, according to Proposition 5.2 of \cite{bogdan64},
\begin{equation*}
    a=(\delta+2vt_1)/(c-v).
\end{equation*}
Thus it is sufficient to know the initial trajectories of the system
on the closed interval $[a,0].$

We should think about such trajectories as a {\em postmortem record} of the trajectory
of some particular body from the system.
It is clear that such trajectories would correspond to
a time interval $J$ that on the left is closed and on the right open or closed,
finite or infinite. In any case it suffices to restrict ourselves to trajectories
defined on intervals of the form $J=[a,b)$ closed on the left and open on the
right. The left end $a$ of such time interval will be called a {\bf point of significance.}
Any time $t_1$ inside of the interval will be called a {\bf stopping time.}

Thus, if our trajectory $t\mapsto r_2(t)$ is of class $C^2,$ from continuity of
the velocity $w(t)=\dot{r}_2(t)$ and of acceleration $\dot{w}(t)$ on the closed
interval $[a,t_1]$ follows that the following two functions
\begin{equation}\label{q and A boundes}
    \begin{split}
    q(t_1)&=\sup\set{|w(u)|:\ u\le t_1, u\in J}<c,\\
    A(t_1)&=\sup\set{|\dot{w}(u)|:\ u\le t_1, u\in J}<\infty,\\
\end{split}
\end{equation}
are well defined for all stopping times $t_1\in J,$ since the supremum of a continuous
function on a closed bounded interval is attained at some point of that interval.

For the sake of mathematical simplicity we shall consider only trajectories defined on
the entire interval $(-\infty,\infty)=R.$

\begin{defin}[Admissible trajectory]
\label{admissible trajectory}
Assume that we are given a path of a point mass $m_0$ that in some Lorentzian frame
has a representation in the form $y=(r_2(t),t),$ where the
function $r_2(t)$ is from $R$ into $R^3$ and it
has continuous derivatives up to order 3 and that for any stoping time $t_1\in J$
the kinetic energy and the acceleration $\ddot{r}_2(t)$ are bounded on the
interval $(-\infty,t_1\rangle.$
We shall say that such a function $r_2(t)$ represents an {\bf
admissible trajectory.}
\end{defin}


\begin{pro}[Kinetic energy bound and velocity bound]
Assume that a body having rest mass $m_0$ moves along a trajectory $r_2:  R  \into R^3.$
Let $c$ denote the speed of light.
For any nonnegative function $k:R\into R$ define function $q:R \into R$
by the formula
\begin{equation}\label{formula for q(t)}
    q(t)=\sqrt{1-\frac{1}{(1+k(t)/(m_0c^2))^2}}\fa t\in   R  .
\end{equation}

Then for any $t\in   R  $ the following two conditions are equivalent
\begin{itemize}
    \item The kinetic energy of the body $m_0$ on the
interval $(-\infty,t\rangle$ is bounded by $k(t).$
    \item The velocity $|v|$ of the body $m_0$ on the
interval $(-\infty,t\rangle$ is bounded by $c\,q(t).$
\end{itemize}
\end{pro}
\bigskip

\bp
    From Einstein's formula \cite{einstein2a}, p. 22,
    the kinetic energy of mass $m_0$ moving with the velocity $v$
    is given by the formula
    \begin{equation*}
        m_0c^2\left(\frac{1}{\sqrt{1-|v|^2/c^2}}-1\right).
    \end{equation*}
    Thus the condition
    \begin{equation*}
        m_0c^2\left(\frac{1}{\sqrt{1-|v(u)|^2/c^2}}-1\right)
        \le k(t)\fa u\le t,\,u\in   R
    \end{equation*}
    is equivalent to the condition
    \begin{equation*}
        |v(u)|\le c\,q(t)\fa u\le t,\,u\in   R  .
    \end{equation*}
    This completes the proof.
\ep
\bigskip

Notice that in the above proposition the quantity $q(t)<1$ for all $t\in R.$
\bigskip


\begin{thm}[Admissible trajectory is relativistic]
The notion of an admissible trajectory does not
depend on the Lorentzian frame, that is if we have two Lorentzian frames $S$ and $S'$
moving with respect to each other with a constant velocity and the path of
the point mass in the frame $S$ forms an admissible trajectory then in the frame
$S'$ the path will form an admissible trajectory as well.
\end{thm}

\bp
    Assume that we have two Lorentzian frames $S$ and $S'.$ Assume that
    the frame $S'$ moves away from frame $S$ with constant velocity $u.$
    Assume that $t\mapsto r_2(t)$ represents an admissible trajectory
    in the frame $S$ and a body with rest mass $m_0$ is moving along the
    trajectory.

    Without loss of generality we may assume
    that the frames $S$ and $S'$ are oriented so that the  transformation
    of the coordinates $y=(r,t)$ from $S$ to $S'$ is given by the formulas
    \begin{equation*}
    \begin{split}
    y'_1&=y_1\\
    y'_2&=y_2\\
    y'_3&=\gamma\,(y_3-uy_4)\\
    y'_4&=\gamma\,(y_4-uy_3)\\
    \end{split}
    \end{equation*}
    where $\gamma=(1-u^2)^{-1/2}$ and $y_4$ and $y'_4$ denote time in
    the respective frames. As before $c=1.$

    First of all notice that the time interval $(-\infty,\infty)$ maps
    onto itself from frame $S$ into $S'.$ Indeed we have
    \begin{equation*}
    \frac{dy'_4}{dy_4}=\gamma\ \frac{dy_4-u\,dy_3}{dy_4}
    =\gamma\,(1-uv)\ge\gamma\ (1-|u|)>0\fa t=y_4\in R.
    \end{equation*}
    Define function $g$ by the formula
    \begin{equation*}
    g(t)=y'_4(y_4)\fa t=y_4\in R.
    \end{equation*}
    From Cauchy's mean value theorem we have,
    for some intermediate point $\theta$ lying between 0 and t,
    \begin{equation*}
    g(t)-g(0)=tg'(\theta)\ge t\,\gamma\ (1-|u|)\fa t>0.
    \end{equation*}
    Thus $y_4=g(t)\into \infty$ if $t\into \infty.$
    Similarly
    \begin{equation*}
    g(t)-g(0)=tg'(\theta)\le t\,\gamma\ (1-|u|)\fa t<0.
    \end{equation*}
    Thus $y_4=g(t)\into -\infty$ if $t\into -\infty.$
    Since any continuous function maps an interval onto an interval
    the function $g$ maps $R$ onto $R.$

    Now introduce a function $f$ by the formula
    \begin{equation*}
        f(w)=\left(\frac{1}{\sqrt{1-w^2}}-1\right)\fa w\ge 0.
    \end{equation*}
    Notice that the function $f$ is nondecreasing and the kinetic
    energy of the mass $m_0$ moving along the trajectory
    can be represented as
    \begin{equation*}
    m_0f(|v|)
    \end{equation*}
    where $$v=\frac{dy_3}{dy_4}=\dot{r}_2$$
    is the  velocity of the body in the frame $S.$

    The velocity of the body in frame $S'$ is given by
    \begin{equation*}
    v'=\frac{dy'_3}{dy'_4}=\frac{dy_3-u\,dy_4}{dy_4-u\,dy_3}=\frac{v-u}{1-uv}.
    \end{equation*}
    Thus we have the estimate
    \begin{equation*}
    |v'|\le \frac{|v|+|u|}{1-|u|}\le \frac{q(t)+|u|}{1-|u|}
    \end{equation*}
    for all times in the initial interval $(-\infty, t\rangle.$
    The quantity  $q(t)$ denotes the velocity bound on the initial interval.
    Thus the velocity $v'$ is bounded on every initial interval $(-\infty,t'\rangle$
    in the frame $S'.$ Therefore its kinetic energy is bounded on every
    initial interval.

    Now let us consider the acceleration in the frame $S'.$ It can be
    expressed as
    \begin{equation*}
    \frac{dv'}{dy'_4}=\frac{\dot{v}(1-u^2)}{\gamma(1-uv)^3}
    \end{equation*}
    in terms of quantities in frame $S.$
    Thus on every initial interval $(-\infty,t\rangle$ we have the estimate
    \begin{equation*}
    \left|\frac{dv'}{dy'_4}\right|\le \frac{A(t)(1-u^2)}{\gamma(1-|u|)^3}
    \end{equation*}
    where $A(t)$ is the bound on the acceleration in the initial
    time interval $(-\infty,t\rangle$ in the frame $S.$
    Hence the acceleration in the frame $S'$ is bounded on every
    initial time interval $(-\infty,t'\rangle.$

    Therefore the trajectory of the moving point mass
    in the frame $S'$ forms an admissible trajectory.
\ep

\section{Retarded time field}
\bigskip

Now consider any point $(r_1,t)$ in a fixed Lorentzian frame and let $(r_2(\tau),\tau)$
denote a point on the path of the point mass with the property that a wave,
emitted from the trajectory at time $\tau,$ and travelling withe the speed $c,$
will arrive at position $r_1$ at time $t.$

The time $\tau$ is called the {\bf retarded time.} It must satisfy the relation
\begin{equation*}
    |r_1-r_2(\tau)|^2-(t-\tau)^2=0,
\end{equation*}
which is preserved under Lorentzian transformations.

The following theorem establishes that the retarded time is
well defined as a function of the variables $(r_1,t)\in R^3\times R.$

\begin{thm}[The retarded time $\tau$ is unique and forms a continuous function]
\label{retarded time is unique}
Assume that we are given in a Lorentzian frame an
admissible trajectory $t\mapsto r_2(t).$
Then  for any point $r_1\in R^3$ and any time $t\in R$
there exists a unique number $\tau\le t$ such that
$$\tau=t-|r_1-r_2(\tau)|.$$

Moreover the map $(r_1,t)\mapsto \tau$ represents a locally Lipschitzian function on the
space $R^3\times R.$ Thus $\tau(r_1,t)$
is continuous on $R^3\times R.$
\end{thm}

\bp
    For fixed $r_1\in R^3$ and $t\in R$ introduce a function $f$
    by the formula
    \begin{equation*}
        f(s)=t-|r_1-r_2(s)|\fa s\le t.
    \end{equation*}
    The function $f$ is well defined and maps the closed interval $(-\infty,t\rangle$
    into itself. The function represents a contraction. Indeed
    \begin{equation}\label{tau is a contraction}
    \begin{split}
        |f(s)-f(\tilde{s})|&=\left|(t-|r_1-r_2(s)|)-(t-|r_1-r_2(\tilde{s})|)\right|
        \le |r_2(s)-r_2(\tilde{s})|\\
        &=|\int_{\tilde{s}}^s v(x)\,dx|\le v_1|s-\tilde{s}|\fa s,\tilde{s}\le t,
    \end{split}
    \end{equation}
    where $v_1=q(t)<c=1$ is the velocity bound corresponding to stopping time $t.$
    Therefore by Banach's contraction mapping theorem there exists one and only one
    solution of the equation $\tau=f(\tau).$

    Thus the map $(r_1,t)\mapsto \tau$ is well defined in our
    Lorentzian frame for all points  $(r_1,t)\in R^3\times R.$

    To prove that the function $\tau$ is locally Lipschitzian
    it suffices to prove that it is Lipschitzian on every open set
    of the form $R^3\times(-\infty,t_1).$  To this end take any two points $(r_1,t)$
    and $(\tilde{r}_1,\tilde{t})$ from the domain of $\tau$
    such that $t,\tilde{t}<t_1.$ Let $v_1<1$ denote the velocity bound corresponding
    to our trajectory on the interval $(-\infty,t_1).$

    To avoid unnecessarily
    complex notation denote by $\tau$ and $\tilde{\tau}$ the retarded times
    corresponding to the points
    $(r_1,t)$ and $(\tilde{r}_1,\tilde{t})$ respectively.
    We have
    \begin{equation*}
     \begin{split}
        |\tau -\tilde{\tau}|&=
        |f(\tau) -f(\tilde{\tau})|=
        \big|(t-|r_1-r_2(\tau)|)-(\tilde{t}-|\tilde{r}_1-r_2(\tilde{\tau})|)\big|\\
            & \le|t-\tilde{t}|+|r_1-\tilde{r}_1|+|r_2(\tau)-r_2(\tilde{\tau})|\\
            & =|t-\tilde{t}|+|r_1-\tilde{r}_1|+|\int_\tau^{\tilde{\tau}}\dot{r}_2(u)\,du|\\
            & \le|t-\tilde{t}|+|r_1-\tilde{r}_1|+v_1|\tau -\tilde{\tau}|.\\
     \end{split}
    \end{equation*}
    Taking the last term in  the above inequality onto the left side and dividing
    by $(1-v_1)$ both sides of the obtained inequality we get
    \begin{equation*}
        |\tau(r_1,t)-\tau(\tilde{r}_1,\tilde{t})|\le
        \frac{1}{1-v_1}(|t-\tilde{t}|+|r_1-\tilde{r}_1|)\fa
        (r_1,t),(\tilde{r}_1,\tilde{t})\in R^3\times (-\infty,t_1)
    \end{equation*}
    Thus the function $\tau$ is continuous on the entire space $R^3\times R.$
\ep

For a proof of Banach's contraction mapping theorem see,
for instance, Loomis and Sternberg
\cite{loomis} page 229.
\bigskip

\begin{thm}[An explicit formula for the retarded time function $\tau$]
\label{explicit formula for retarded time}
Assume that we are given in a Lorentzian frame an
admissible trajectory $t\mapsto r_2(t).$

Take any stopping time $t_1\in R$
and let $v_1=q(t_1)<c$ denote the corresponding velocity bound for $t\le t_1$
where the function $q$ is given by the formula \ref{formula for q(t)}.


Put $s_0(r_1,t)=0$ and define recursively the sequence
\begin{equation*}
    s_n(r_1,t)=f(s_{n-1}(r_1,t))\fa n=1,2,3,\dots;\text{ and }r_1\in R^3,\ t\le t_1,
\end{equation*}
where $f(s)=t-|r_1-r_2(t-s)|$ for all $s\le t.$

The retarded function $\tau$ is given by the formula
\begin{equation*}
    \tau(r_1,t)=\lim_n s_n(r_1,t) \fa  r_1\in R^3\text{ and }t\in R.
\end{equation*}
Moreover we have the following convenient estimate for the rate of convergence
\begin{equation*}
    |\tau(r_1,t)-s_n(r_1,t)|\le \frac{v_1^n}{1-v_1}\big|t-|r_1-r_2(t)|\big|\fa r_1\in R^3\text{ and }t\le t_1.
\end{equation*}
\end{thm}

\bp
The proof follows from Theorem 4.7 of Bogdan \cite{bogdan64} or
Theorem 9.1 on page 229
of Loomis and Sternberg \cite{loomis}.
\ep

\bigskip

\section{The fundamental fields associated with\\ an admissible trajectory}
\bigskip

Now define  the delay function $T(r_1,t)=t-\tau(r_1,t)$ and notice that it satisfies
the equation
\begin{equation}\label{Equ for T}
    T=|r_1-r_2(t-T)|\fa t\in R\text{ and }r_1\in R^3.
\end{equation}

Since the function $T$ as difference of two continuous functions is continuous
the set
\begin{equation*}
 G=\set{(r_1,t)\in R^3\times R:\ T(r_1,t)>0}=T^{-1}(0,\infty)
\end{equation*}
as an inverse image of an open set by means of a continuous function is itself open.
The set $G$ consists of points that do not lie on the trajectory.

By assumption the trajectory $r_2$ has continuous derivatives $\dot{r}_2(t)=w(t)$
and $\dot{w}(t),$
so we can define the vector fields
$$r_{12}(r_1,t)=r_1-r_2(\tau(r_1,t)),\quad v(r_1,t)=\dot{r}_2(\tau(r_1,t))
,\quad a(r_1,t)=\dot{w}(\tau(r_1,t))$$
for all $(r_1,t)\in R^4.$
Introduce the unit vector field $e=r_{12}/T$ and
fields $u$ and $z$ by the formulas
$$
    u=\frac{1}{T}\qtext{and}z=\frac{1}{(1- \dotp{e}{v} )}\qtext{on}G.
$$
In the above $\dotp{e}{v}$ denotes the dot product of the vectors $e$ and $v.$
Since $|\dotp{e}{v}|\le |v|<c=1$ the vector field $z$ is well defined on the
set $G.$


\begin{defin}[Fundamental fields]
\label{Fundamental fields}
Assume that we are given in a Lorentzian frame an
admissible trajectory $t\mapsto r_2(t).$

Define the time derivative $w(t)=\dot{r}_2(t).$
The fields given by the formulas
\begin{equation*}
    \tau,\ T,\ r_{12}=r_1-r_2\circ \tau,\ v=w\circ \tau,\ a=\dot{w}\circ \tau\fa (r_1,t)\in G
\end{equation*}
and
\begin{equation*}
    u=1/T,\quad  e=u\, r_{12},\qtext{and} z\quad\fa (r_1,t)\in G
\end{equation*}
will be called the {\bf fundamental fields}
associated with the trajectory $r_2(t).$
The operation $\circ$ denotes here the composition of functions.
\end{defin}
\bigskip

The fundamental fields are continuous on their respective
domains.
This follows from the fact that composition of continuous functions yields
a continuous function. Thus all of them, for sure,
are continuous on the open set $G$ of
points that do not lie on the trajectory.

Analogous fields defined by similar formulas on an open set $G\subset R^4$
appear in the problems involving
plasma flows \cite{bogdan71} and \cite{bogdan72}, or more generally flows of matter.

\bigskip

We would like to stress here that the fundamental fields depend on the Lorentzian
frame, in which we consider the trajectory.
It is important to find expressions involving fundamental fields that
yield fields invariant under Lorentzian transformations.

Lorentz and Einstein \cite{einstein2a}, Part II, section 6, established that
fields satisfying homogeneous Maxwell equations are
invariant under Lorentzian transformations.
\bigskip

In the paper Bogdan \cite{bogdan70} our main goal was
to prove that fields given by amended Feynman formulas and
fields obtained from Li\'{e}nard-Wiechert potentials satisfy
Maxwell equations.

We have proved this fact by showing that these fields
are representable by means of fundamental fields and using the formulas
for partial derivatives of the fundamental fields proved that
such fields generate fields satisfying homogeneous Maxwell equations.

The following theorem represents the main pillar of the argument.
For the sake of completeness
we shall present the proof of the theorem in its entirety.
\bigskip

Introduce operators $D=\frac{\partial}{\partial t}$ and
$D_i=\frac{\partial}{\partial x_{i}}$ for $i=1,2,3$ and
$\nabla=(D_1,D_2,D_3).$

Observe that $\delta_i$ in the following formulas denotes the i-th
unit vector of the standard base in $R^3$ that is
$\delta_1=(1,0,0),$ $\delta_2=(0,1,0),$ $\delta_3=(0,0,1).$


\begin{thm}[Partial derivatives of fundamental fields]
Assume that in some Lorentzian frame
we are given an   admissible trajectory $t\mapsto r_2(t).$
Define the time derivative $w(t)=\dot{r}_2(t).$
For partial derivatives with respect to coordinates of the vector $r_1$
we have the following identities on the set $G$
\begin{eqnarray}
\label{DiT}       D_iT&=&ze_i  \qtext{where} z=(1- \dotp{e}{v} )^{-1} ,\quad v=\dot{r}_2\circ\tau,\\
\label{Diu}       D_iu&=&-zu^2e_i  ,   \\
\label{Div}          D_iv&=&-e_iza \qtext{where} a=\dot{w}\circ\tau,\\
\label{Di tau}    D_i\tau&=&-ze_i,\\
\label{Die}         D_ie&=& -uze_ie+u\delta_i+uze_iv \qtext{where} \delta_i=(\delta_{ij}),\\
\label{Diz}          D_i z &=& -z^3e_i \langle e,a \rangle -uz^3e_i+ uz^2e_i+uz^2v_i+uz^3e_i \langle v,v \rangle
\end{eqnarray}
and for the partial derivative with respect to time we have
\begin{eqnarray}
\label{DT}    DT&=&1-z,\\
\label{Du}    Du&=&zu^2-u^2,\\
\label{D tau}    D\tau&=&z,\\
\label{Dv}    Dv&=&za  \qtext{where} a=\dot{w}\circ\tau,\\
\label{De}    De&=&-u e+ u z e-u z v,\\
\label{Dz}    Dz&=&uz-2uz^2 +z^3 \langle e,a \rangle +uz^3-uz^3 \langle v,v \rangle .
\end{eqnarray}
Since the expression on the right side of each formula represents
a continuous function, the fundamental fields are at least of class $C^1$ on the set $G.$

Moreover if the trajectory is of class $C^\infty$ then one can easily deduce from the above
formulas that the fundamental
fields are of class $C^\infty$ on $G$ as well.
\end{thm}

\bigskip

\bp
Proof of formula (\ref{DiT}):
Applying the operator $D_i$ to both sides of equation (\ref{Equ for T}) we get
\begin{equation*}
\begin{split}
    D_iT&=\langle e,D_i(r_1-r_2) \rangle
    =\langle e,(\delta_{ij}) \rangle -\langle e,w(\tau) \rangle D_i\tau\\
        =&e_i-\langle e,w(\tau) \rangle (-D_iT)=e_i+ \langle e,v \rangle D_iT
\end{split}
\end{equation*}
yielding formula (\ref{DiT}).

Formula (\ref{Diu}) 
follows from formula (\ref{DiT}).

Proof of formula (\ref{Div}): $$D_iv=aD_i\tau=a(-D_iT)=-e_iza$$

Proof of formula (\ref{Di tau}): $$D_i\tau=D_i(t-T)=-D_iT=-ze_i$$

Proof of formula (\ref{Die}):
\begin{equation}
\begin{split}
    D_ie&=D_i[u(r_1-r_2)]=[D_iu]r_{12}+u[D_ir_1-D_ir_2]\\
    &=[D_iu]u^{-1}e+u[D_ir_1-D_ir_2]\\
    &=[-zu^2e_i]u^{-1}e+u\delta_i-u[D_i\tau]v\\
    &=-zue_ie+u\delta_i-u[-ze_i]v\\
    &=-zue_ie+u\delta_i+uze_iv\\
    &=-uze_ie+u\delta_i+uze_iv
\end{split}
\end{equation}

Proof of formula (\ref{Diz}):
\begin{equation}
\begin{split}
    D_i z&= D_i(1- \langle e,v \rangle )^{-1}=(-1)z^2(-D_i \langle e,v \rangle )\\
    &=z^2[ \langle e,D_iv \rangle + \langle v,-uze_ie+u\delta_i+uze_iv \rangle ]\\
    &=z^2[ \langle e,-e_iza \rangle + \langle v,-zue_ie+u\delta_i+uze_iv \rangle ]\\
    &=-z^3e_i \langle e,a \rangle -uz^3e_i
    \langle v,e \rangle +uz^2v_i+uz^3e_i \langle v,v \rangle \\
    &=-z^3e_i \langle e,a \rangle -uz^3e_i[1-z^{-1}]
    +uz^2v_i+uz^3e_i \langle v,v \rangle \\
    &=-z^3e_i \langle e,a \rangle -uz^3e_i+ uz^2e_i
    +uz^2v_i+uz^3e_i \langle v,v \rangle \\
\end{split}
\end{equation}

\bigskip

Proof of formula (\ref{DT}): Applying the operator
$D$ to both sides of the equation (\ref{Equ for T})
we get
\begin{equation*}
\begin{split}
    DT&=D|r_{12}|=(e,D[r_{12}])=-(e,Dr_2)\\
    &=- \langle e,v \rangle D\tau=- \langle e,v \rangle (1-DT)
    =- \langle e,v \rangle + \langle e,v \rangle DT.
\end{split}
\end{equation*}
The above yields
\begin{equation*}
\begin{split}
    DT&=\frac{- \langle e,v \rangle }{1- \langle e,v \rangle }=1-z.
\end{split}
\end{equation*}

Proof of formula (\ref{Du}): $$Du=DT^{-1}
=(-1)T^{-2}DT=(-1)u^2(1-z)=zu^2-u^2.$$

Proof of formula (\ref{D tau}): $$D\tau=D(t-T)=1-DT=1-(1-z)=z.$$

Proof of formula (\ref{Dv}): $$Dv=aD\tau=za.$$

Proof of formula (\ref{De}):
\begin{equation*}
\begin{split}
    De&=D[ur_{12}]=[Du]r_{12}+u[Dr_{12}]\\
    &=[zu^2-u^2]u^{-1}e-u[Dr_2]\\
    &=[zu-u]e-u[(D\tau)v]\\
    &=zue-ue-uzv=-ue+uze-uzv.
\end{split}
\end{equation*}

Proof of formula (\ref{Dz}):
\begin{equation*}
\begin{split}
    Dz&=D(1- \langle e,v \rangle )^{-1}
        =(1- \langle e,v \rangle )^{-2}D \langle e,v \rangle\\
    &=z^2 \langle De,v \rangle +z^2 \langle e,Dv \rangle \\
    &=z^2 \langle [uze-ue-uzv],v \rangle +z^2 \langle e,za \rangle \\
    &=uz^3 \langle e,v \rangle -uz^2 \langle e,v \rangle
    -uz^3 \langle v,v \rangle +z^3 \langle e,a \rangle \\
    &=uz^3[1-z^{-1}]-uz^2[1-z^{-1}]-uz^3 \langle v,v \rangle
    +z^3 \langle e,a \rangle \\
    &=uz^3-uz^2 -uz^2+uz-uz^3 \langle v,v \rangle +z^3 \langle e,a \rangle \\
    &=uz^3-2uz^2 +uz-uz^3 \langle v,v \rangle +z^3 \langle e,a \rangle \\
    &=uz-2uz^2 +z^3 \langle e,a \rangle +uz^3-uz^3 \langle v,v \rangle \\
\end{split}
\end{equation*}
\ep

\bigskip

\section{Bogdan-Feynman Theorem for a moving point mass}
\bigskip

We shall present here Bogdan-Feynman theorem containing amended Feynman Law
as its consequence. The proof of this theorem is based on the differentiation
formulas of the fundamental fields and is presented in Bogdan \cite{bogdan70}.

Partial derivatives with respect to coordinates of $r_1$
are denoted by $D_1,\,D_2,\,D_3$ and with respect to time just by $D.$
The gradient differential operator is denoted by $\nabla=(D_1,D_2,D_3)$
and the D'Alembertian operator by $\Box^2=\nabla^2-D^2.$

\begin{thm}[Bogdan-Feynman Theorem]
\label{Bogdan-Feynman Theorem}
Assume that in a given Lorentzian frame the map $t\mapsto r_2(t)$
from $R$ to $R^3$ represents an
admissible trajectory of class $C^3.$ 
Assume that $G$ denotes the open set of points
that do not lie on the trajectory.
All the following field equations are satisfied on the entire set $G.$

Consider the pair of fields $E$ and $B$ over the set $G$
given by the formulas
\begin{equation*}
     E=u^2e+u^{-1}D(u^2e)+D^2e\qtext{and}B=e\times E
\end{equation*}
where $u$ and $e$ represent fundamental fields (\ref{Fundamental fields})
associated with the trajectory $r_2(t).$

Then this pair of fields will satisfy
the following homogenous system of Maxwell equations
\begin{equation*}
    \begin{split}
    &\nabla\times E=-DB,\quad \nabla\cdot E=0,\\
    &\nabla\times B=+DE,\quad \nabla\cdot B=0,\\
    \end{split}
\end{equation*}
and the homogenous wave equations
\begin{equation*}
    \Box^2 E=0,\qquad \Box^2 B=0.
\end{equation*}

Moreover Li\'{e}nard-Wiechert potentials, expressed in terms of the
fundamental fields as  $A=uzv$ and $\phi=uz,$ satisfy the homogenous
system of wave equations with Lorentz gauge formula
\begin{equation*}
    \Box^2 A=0,\quad \Box^2\phi=0,\quad \nabla\cdot A+D\phi=0
\end{equation*}
and generate the fields $E$ and $B$ by the formulas

\begin{equation*}
\begin{split}
    E&=-\nabla \phi-DA\qtext{and}B=\nabla\times A,\\
\end{split}
\end{equation*}

Finally we have the following explicit formula for the field $E$
in terms of the fundamental fields
\begin{equation*}\label{formula4F-b}
\begin{split}
    E&= -uz^2a   +uz^3 \langle e,a \rangle e -uz^3 \langle e,a \rangle v \\
    &\quad  +u^2z^3e-u^2z^3 \langle v,v \rangle e-u^2z^3v+u^2z^3 \langle v,v \rangle v.\\
\end{split}
\end{equation*}
\end{thm}
\bigskip

As a consequence of the above theorem the components of the quantities $E,$ $B,$
$A,$ and $\phi$ propagate in the Lorentzian frame with velocity of light $c.$
\bigskip


\bigskip
\section{Representing arbitrary fields by means\\ of fundamental fields}
\bigskip

Assume that in Lorentzian frame $S$ we have
an admissible trajectory $t\mapsto r_2(t)$ of a point mass.

We will prove that the map $\phi:(t,r)\mapsto (\tau,T,e)$ from the open set
\begin{equation}\label{set G}
    G=\set{(r,t)\in R^3\times R:\ T(r,t)>0}
\end{equation}
consisting of points that do not lie on the trajectory into the set
\begin{equation}\label{image of G}
    G_1=R\times J\times S_2,
\end{equation}
where
\begin{equation*}
    J=(0,\infty)\qtext{and}S_2=\set{r\in R^3:\ |r|=1},
\end{equation*}
is one to one and onto and represents a diffeomorphism, that is both $\phi$ and its
inverse $\phi^{-1}$ are differentiable.

\begin{thm}[Map $\phi$ is a diffeomorphism onto]
Let $t\mapsto r_2(t)$ represent an admissible trajectory of a point mass in
some Lorentzian frame $S.$

The map $\phi$ considered on the open set $G$ of points not lying on the trajectory
$r_2$ represents a diffeomorphism of the set $G$ onto the set $G_1$ of class $C^1.$

Moreover if the trajectory $r_2$ is of class $C^\infty$ then the diffeomorphism is
also of class $C^\infty.$
\end{thm}
\bigskip

\bp
    Take any point $(r,t)\in G$ and notice that by the definition of the fields
    $\tau$ and $T$ and $e$ we have $\tau(r,t)\in R,$ $T(r,t)\in J,$ and $e(r,t)$ is
    a unite vector that is $e(r,t)\in S_2.$
    Thus the map $\phi$ is from $G$ into $G_1.$

    Now take any point $(\tau,T,e)\in G_1$ and define the map $\psi$ by the formula
    \begin{equation*}
        \psi:(\tau,T,e)\mapsto (r,t)=(r_2(\tau)+Te,\tau+T).
    \end{equation*}
    Plainly the map $\psi$ is from $G_1$ into $G.$ Moreover the maps $\phi$
    and $\psi$ are inverse to each other, thus $\phi$ establishes a one to
    one correspondence between points of the set $G$ and the set $G_1.$

    The sphere $S_2$ represents a differentiable manifold. We can define
    a system of six maps, two for each coordinate axis, to provide
    a complete atlas for the entire sphere. Thus for instance for the
    upper hemisphere we have the map over the set
    $$\set{(e_1,e_2)\in R^2:\ e_1^2+e_2^2<1},$$
    which provides a representation for the upper hemisphere
    \begin{equation*}
         (e_1,e_2,(1-e_1^2-e_2^2)^{1/2}).
    \end{equation*}

    It follows from the symmetry of the problem, that to prove that the transformation
    $\phi$ forms a diffeomorphism, it is sufficient to show that
    the Jacobian
    \begin{equation*}
         L=\frac{\partial (\tau,T,e_1,e_2)}{\partial (t,x_1,x_2,x_3)},
    \end{equation*}
    where $x_i$ represent coordinates of the vector $r\in R^3,$
    is nonzero. For reference see Cartan \cite{cartan}, page 51, Corollary 4.2.2.

    We have
\begin{equation}\label{Jacobian}
    L=\begin{vmatrix}
      \frac{\partial \tau}{\partial t} & \frac{\partial T}{\partial t} &
      \frac{\partial e_1}{\partial t} & \frac{\partial e_2}{\partial t} \\ \\
      \frac{\partial \tau}{\partial x_1} & \frac{\partial T}{\partial x_1} &
      \frac{\partial e_1}{\partial x_1} & \frac{\partial e_2}{\partial x_1} \\ \\
      \frac{\partial \tau}{\partial x_2} & \frac{\partial T}{\partial x_2} &
      \frac{\partial e_1}{\partial x_2} & \frac{\partial e_2}{\partial x_2} \\ \\
      \frac{\partial \tau}{\partial x_3} & \frac{\partial T}{\partial x_3} &
      \frac{\partial e_1}{\partial x_3} & \frac{\partial e_2}{\partial x_3} \\
    \end{vmatrix}
\end{equation}

    Adding the second column to the first one and expanding the resulting determinant
    with respect to the first column we get
\begin{equation}\label{Jacobian2}
    L=
    \begin{vmatrix}
       \frac{\partial T}{\partial x_1} &
      \frac{\partial e_1}{\partial x_1} & \frac{\partial e_2}{\partial x_1} \\ \\
     \frac{\partial T}{\partial x_2} &
      \frac{\partial e_1}{\partial x_2} & \frac{\partial e_2}{\partial x_2} \\ \\
      \frac{\partial T}{\partial x_3} &
      \frac{\partial e_1}{\partial x_3} & \frac{\partial e_2}{\partial x_3} \\
    \end{vmatrix}
    =
    \begin{vmatrix}
      {ze_1} & D_1 e_1 & D_1 e_2 \\ \\
      {ze_2} & D_2 e_1 & D_2 e_2 \\ \\
      {ze_3} & D_3 e_1 & D_3 e_2 \\
    \end{vmatrix}
    =zu^2
    \begin{vmatrix}
      {e_1} & 1 & 0 \\ \\
      {e_2} & 0 & 1 \\ \\
      {e_3} & 0 & 0 \\
    \end{vmatrix}=zu^2e_3\neq0
\end{equation}

\ep

\bigskip

\begin{cor}[Representation of arbitrary field over $G$]
Any field over the set $G$ can be represented as a function of the
following three fields $\tau,$ $T,$ and $e.$
\end{cor}

The above corollary justifies the terminology that we used naming the fields
$\tau,$ $T,$ $e,$ $v,$ $a,$ $u,$ $z$
{\em fundamental.} In spite of generality of the above result the representations of
some important fields like $v,$ $a,$ and for instance $z$ are not simple.

The set $G_1$ has a simple structure which should be explored. The space $R$
considered as a group under addition is equipped with invariant under translations
integral, the Lebesgue integral. The set $J$ considered as a group under multiplication
is isomorphic to the additive group $R,$ and the sphere $S_2$ has an integral that is
invariant under linear isometries of $R^3.$ So one can develop wavelet structure
like in Kaiser \cite{kaiser1}.

On top of this $G_1$ can be considered as a differentiable manifold of class $C^\infty$
so the theory of generalized functions similar to theories of Schwartz \cite{schwartz}
and Gelfand \cite{gelfand1} can be developed.


\end{document}